\documentclass[usegraphicx,usenatbib,a4paper]{mn2e}
\usepackage{aas_macros,graphicx,times,multirow}
\usepackage{amssymb}
\usepackage{float}
\usepackage[T1]{fontenc}
\usepackage{aecompl}

\title[Two new Intermediate Polars]
{\aa\ and \bb: two new hard X-ray selected magnetic cataclysmic variables identified with \XMM}
\author[F. Bernardini et al.]                                                    
{F.~Bernardini,$^{1,2}$\thanks{E-mail: bernardini@nyu.edu} 
D.~de Martino,$^{2}$ 
K.~Mukai,$^{3,4}$
G. Israel,$^{5}$
M.~Falanga,$^{6,7}$
G.~Ramsay,$^{8}$
\newauthor
N.~Masetti$^{9,10}$\\
$^1$ New York University Abu Dhabi, Saadiyat Island, Abu Dhabi, 129188, United Arab Emirates\\
$^2$ INAF $-$ Osservatorio Astronomico di Capodimonte, Salita Moiariello 16, I-80131 Napoli, Italy\\
$^3$ CRESST and X-Ray Astrophysics Laboratory, NASA Goddard Space Flight Center, Greenbelt, MD 20771, USA\\
$^4$ Department of Physics, University of Maryland, Baltimore County, 1000 Hilltop Circle, Baltimore, MD 21250, USA\\
$^5$ INAF - Osservatorio Astronomico di Roma, via Frascati 33, I-00040 Monteporzio Catone, Roma, Italy\\
$^6$ International Space Science Institute (ISSI), Hallerstrasse 6, CH-3012 Bern, Switzerland\\
$^7$ International Space Science Institute in Beijing, No. 1 Nan Er Tiao, Zhong Guan Cun, Beijing 100190, China\\
$^8$ Armagh Observatory, College Hill, Armagh BT61 9DG, UK\\
$^9$ INAF - Istituto di Astrofisica Spaziale e Fisica Cosmica, Via Gobetti 101, I-40129, Bologna, Italy \\
$^{10}$ Departamento de Ciencias F{\'i}sicas, Universidad Andr{\'e}s Bello, Fern{\'a}ndez Concha 700, Las Condes, Santiago, Chile} 

\date{}
\pagerange{\pageref{firstpage}--\pageref{lastpage}}
\def\Swift{{\em Swift}}

\def\XMM{{\em XMM-Newton}}

\def\ergscm{$\rm erg\,cm^{-2}\,s^{-1}$}
\def\INT{{\em INTEGRAL}\,}

\def\nh{\hbox{$N_{\rm H}$}}
\def\aa{Swift\,J0525.6+2416}
\def\bb{IGR\,J04571+4527}

\begin{document}

\label{firstpage}

\maketitle

\begin{abstract}

\bb\ and \aa\ are two hard X-ray sources detected in the \Swift/BAT and INTEGRAL/IBIS surveys. They were proposed to be magnetic cataclysmic variables of the Intermediate Polar (IP) type, based on optical spectroscopy. \bb\ also showed a 1218 s optical periodicity, suggestive of the rotational period of a white dwarf, further pointing towards an IP classification. We here present detailed X-ray (0.3--10 keV) timing and spectral analysis performed with \XMM, complemented with hard X-ray coverage (15--70 keV) from \Swift/BAT. These are the first high S/N observations in the soft X-ray domain for both sources, allowing us to identify the white dwarf X-ray spin period of \aa\ ($226.28$ s), and \bb\ ($1222.6$ s). A model consisting of multi-temperature optically thin emission with
complex absorption adequately fits the broad-band spectrum of both sources. We estimate a white dwarf mass of about 1.1 and 1.0 M$_{\odot}$ for \bb\ and \aa, respectively. The above characteristics allow us to unambiguously classify both sources as IPs, confirming the high incidence of this subclass among hard X-ray emitting Cataclysmic Variables. 

\end{abstract}


\section{Introduction}

Binary systems containing an accreting white dwarf (WD) from a late-type main sequence or sub-giant star  are known as Cataclysmic Variable stars (CVs). Thanks to the recent hard X-ray surveys carried out by \Swift\ Burst Alert Telescope \citep[BAT;][]{barthelmy} and \INT\ IBIS \citep{ubertini03} instruments above 20 keV, the number of detected magnetic systems (where $B\gtrsim10^{5}$ G) is rapidly increasing \citep[][and references therein]{barlow06,bernardini12,bernardini13,bernardini14}. Magnetic CVs are about 20 per cent of all CVs \citep{ferrario15} and they are further divided in two main groups, depending on the degree of synchronization ($P_{rot=\omega}/P_{orb=\Omega}$). The synchronous systems, called polars, are polarised in the optical and near-IR bands revealing WDs with strong magnetic fields ($B\sim 7-230\times10^{6}$ G). These are able to lock the rotation of the WD with the orbital period. 
Those systems which instead possess an asynchronously rotating WD are called intermediate polars (IPs) and generally do not show optical/nearIR polarisation, suggesting they are weakly magnetised WDs ($B\leq10^{6}$ G).
In magnetic CVs, the accretion flow is magnetically channelled on the WD polar regions. The high field Polars accrete matter
directly through a stream from the Roche lobe overflowing donor star, while the lower field IPs may accrete through a stream or a truncated disc or ring, depending on their degree of asynchronism and their magnetic moment \citep{norton04,norton08}.
The the magnetically confined accretion flow approaches supersonic velocities and a strong stand-off shock ($\sim10-80$ keV) is formed. In the post-shock region matter slows and cools via cyclotron radiation emerging in the optical/near-IR and/or via thermal bremsstrahlung emerging in the X-rays \citep{aizu73,wu94,cropper99}. The dominant cooling channel primarily depends on the WD magnetic field intensity \citep{woelk_beuermann96,fischer_beuermann01}. The post-shock region in the polars mainly cool via cyclotron emission, while IPs are bremmstrhalung dominated systems.
The post-shock emission is intercepted and thermalised by the WD surface, and can appear to a far observer as an optically thick, highly absorbed, spectral component emerging in the soft X-ray ($\sim20-60$ eV) or in the ultraviolet or extreme ultraviolet (UV/EUV). 

Magnetic CVs are debated as important contributors of the X-ray luminosity function at low luminosities ($10^{30}-10^{33}$ erg/s) as inferred from X-ray surveys of the galactic ridge and bulge \citep{sazonov06,revnivtsev08,revnivtsev09,revnivtsev11,hong12a,hong12b}.  
The fact that they are about 6\% of the galactic hard X-ray sources in the Swift/BAT and INTEGRAL/IBIS surveys \citep{barlow06,bird10,cusumano10,baumgartner13} supports this scenarios. IPs represent about 80\% of these hard X-ray CVs, and their number is increasing thanks to follow-ups in the X-rays and optical bands \citep[see e.g.][]{barlow06,anzolin09,bernardini12,bernardini13}.
The characterisation of new sources, especially the faint ones, may allow inferring new aspects of the luminosity function, such as a possible large, but still to-be-discovered, population of low-luminosity and short orbital period IPs \citep[see][]{pretorius14}. It will also allow understanding the role played by main source parameters, such as the WD mass, magnetic field intensity, and the mass accretion rate in shaping the evolution of these systems and their X-ray spectral properties.
X-rays are the best diagnostic bands to identify the true WD spin period, thus allowing a secure classification. Moreover, new properties, such as the presence of a soft blackbody component and of a ionised (warm) absorber have been recently identified in an increasing number of IPs thanks to the unique capabilities of the XMM-Newton satellite \citep[see e.g.][]{mukai01,demartino08,bernardini12}. 

We here present the first high S/N soft X-ray observations, carried out with \XMM, of \aa\ (also known as 1RXS J052523.2+241331 and as WISE J052522.84+24133333.6) and \bb\ (also known as 1RXS J045707.4+452751 and as Swift J0457.1+4528). Based on the characteristics of its optical spectra, \aa\ was proposed to be a magnetic CV or a high accretion rate novalike \citep{torres07,masetti12a}. \cite{ramsay09b} reported the possible detection of an eclipse from this source by using optical spectroscopic data. \bb, instead, was first classified as a non-magnetic CV \citep{masetti10}. Later on, the unambiguous detection of a coherent signal at $1218.7\pm0.5$ s in optical photometric data, interpreted as the WD spin period, together with the detection of a spectroscopic period \citep[4.8 h or 6.2 h, one the alias of the other;][]{thorstensen13}, provided strong arguments on the IP nature of \bb.

We complement the X-ray study of both sources with simultaneous optical photometry acquired with the Optical Monitor \citep{mason01} on board \XMM\ and for \aa\ we also include a re-analysis of the optical spectroscopic data presented in \cite{ramsay09b} and  Wide-field Infrared Survey Explorer (WISE) archival photometry.

\section{Observations and data analysis}
\label{sec:obs}

\subsection{\textit{XMM-Newton} observations}

\aa\ and \bb\ were observed on 21 February 2014 and 13 March 2014, respectively, with the EPIC PN, MOS1, and MOS2 cameras \citep{struder01,turner01} as main instruments. 
The log of the observations is reported in Table \ref{tab:observ}. Data were processed using the \textsc{SAS} version 13.0.0 and the latest calibration files (CCF) available on October 2014.

\begin{table*}
\caption{Summary of main observations parameters for all instruments. Uncertainties are at $1\sigma$ confidence level.}
\begin{center}
\begin{tabular}{cccccccc}
\hline 
Source & Telescope      & OBSID & Instrument & Date                    & UT$_{\rm start}$ & T$_{\rm expo}^{a}$  & Net Source Count Rate\\
                    &               &              &        & yyyy-mm-dd      & hh:mm & (ks)      &    c/s                  \\
\hline
\bb & \emph{XMM-Newton}    &  0721790201  & EPIC-PN$^b$  & 2014-03-13  & 11:42 & 36.1 & $1.424\pm0.008$ \\
     &                      &              & EPIC-MOS1$^c$ & 2014-03-13  & 11:19 & 37.8 & $0.491\pm0.004$     \\ 
     &                      &              & EPIC-MOS2$^c$ & 2014-03-13  & 11:19 & 37.8 & $0.494\pm0.004$     \\  
     &                      &              & OM-B$^d$     & 2014-03-13  & 11:25 & 32.0 & $2.33\pm0.07$ \\
     &         \emph{Swift} & $^{e}$       & BAT &  &	& $8.5\times10^{6}$ & $3.9\pm0.3\times 10^{-5}$ \\ 
\hline              
\aa  & \emph{XMM-Newton}    &  0721790301  & EPIC-PN$^b$  & 2014-02-21  & 07:15 & 30.0 & $1.494\pm0.008$ \\
     &                      &              & EPIC-MOS1$^c$ & 2014-02-21  & 06:52 & 31.7 & $0.574\pm0.04$     \\ 
     &                      &              & EPIC-MOS2$^c$ & 2014-02-21  & 06:52 & 31.7 & $0.577\pm0.04$     \\  
     &                      &              & OM-B      & 2014-02-21  & 06:58 & 26.6 & 4.78$\pm0.03$ \\
     &      \emph{Swift}    & $^{e}$       & BAT &  &	& $7.3\times10^{6}$ & $3.7\pm0.4\times 10^{-5}$ \\ 
\hline
\end{tabular}
\label{tab:observ}
\end{center}
\begin{flushleft}
$^a$ Net exposure times.\\
$^b$ Full frame mode (thin filter applied). \\
$^c$ Large window mode (thin filter applied). \\
$^d$ Fast window mode. The central wavelength of the filter is 4500 \AA. \\ 
$^e$ All available pointings collected during 2004 December to 2010 September are summed together. \\
\end{flushleft}
\end{table*}

\subsubsection{The EPIC data}

We extracted the source photons from a circular region with a radius of 40 arcsec, centered at the source position. Background photons were taken from a nearby empty region of the sky, with a radius of 70 (PN) and 140 (MOSs) arcsec. 
For \aa\ we used the whole data set for both the spectral and the timing analysis (the high background contamination is only a small fraction of the source flux). In the case of \bb, the high background epoch is mainly concentrated at the end of the pointing, consequently we removed this part of the observation. For the timing analysis, instead, we used the whole data set. 

We produced background-subtracted light curves in the ranges $0.3-12$ keV, $0.3-2$ keV, $2-3$ keV, $3-5$ keV and $5-12$ keV (for \aa\ we also used the $0.3-1$ and $1-3$ keV range). The event arrival times were corrected to Solar system barycentre by using the task \textsc{barycen}. 
To study spectral variability with the source rotation we extracted the EPIC spectra at pulse maximum and minimum.

The EPIC spectra were rebinned before fitting with the tool \textsc{specgroup} in order to have a minimum of 50 (PN) and 25 (MOS) counts each bin. The spectral fit were made with \textsc{Xspec} version 12.8.2.
The sources are too faint for a use of the RGS spectra.

\subsubsection{The Optical Monitor photometry}

We run the task \textsc{omfchain} to produce the B-band light curves. 
However, due to offset coordinates, the pointing to \bb\ had the fast window centred on a different star. Thus, no fast timing on this source is available, but magnitudes were obtained from the 2-D images that are acquired along each of the ten $\sim$3200 s long  segments. In these images \bb\ is stable at an average magnitude B$=18.35\pm0.03$ corresponding to a flux of $3\times10^{-16}$ erg cm$^{-2}$ s$^{-1}$ \AA$^{-1}$. It is 1.6 times fainter than when observed in 2009 by \cite{masetti10} and at about the same level than when observed in 2010 by \cite{thorstensen13}. 
\noindent
\aa\ has an average magnitude B$=17.57\pm0.01$ corresponding to a B-band
flux of $6.2\times10^{-16}$ erg cm$^{-2}$ s$^{-1}$ \AA$^{-1}$, which is comparable with what found by \cite{masetti12a}
and reported in the USNO A2.0 catalogue. 

\subsubsection{The optical spectroscopy of \aa}

Spectroscopic observations of SWIFT J0525+2413 were obtained using the
4.2m William Herschel Telescope (WHT) equipped with the Intermediate dispersion
Spectrograph and Imaging System (ISIS) on La Palma. Four spectra,
each with an exposure time of 240 sec, were taken consecutively on the
night of 2008 Oct 6. The cadence was $\sim$9 min with a start-to-end
duration of $\sim$32 min. Conditions on the night were clear, but with
variable seeing. We used the R300B ($\sim$3500--5300 \AA) and R158R ($\sim$5700--10000 \AA) gratings
together with a 0.8 arcsec slit.  A fit to the arc lines gave
$\sigma=$1.9 and $\sigma=$2.2 \AA\ for the blue and red
spectra, respectively.
The data were reduced using the {\sc Starlink}\footnote{The Starlink
  Software Group homepage can be found at
  http://starlink.jach.hawaii.edu/starlink} {\sc Figaro} package. 
The wavelength calibration was obtained using CuAr+CuNe arc lamp exposures taken before the series of spectra on the target. Observations of the spectroscopic standard He3 were used to remove the instrumental response.
Due to the variable seeing during the night some flux could have been lost and therefore we did not calibrate the absolute flux scale.

\subsection{The \Swift\ observations}

Both sources are catalogued in the 70 month Swift/BAT source catalogue (Baumgartner et al. 2013). We downloaded the archival 
eight-channel spectra of the two sources and the response matrix files that are directly available at http://swift.gsfc.nasa.gov/results/bs70mon/.
Due to the low S/N of the BAT spectra above 70k eV we limited the analysis of the broad-band spectra up to this energy.

\section{Results}
\label{sec:results}

\subsection{\bb}

\subsubsection{The X-ray and optical timing analysis}
\label{subsub:04timing}

\begin{figure}
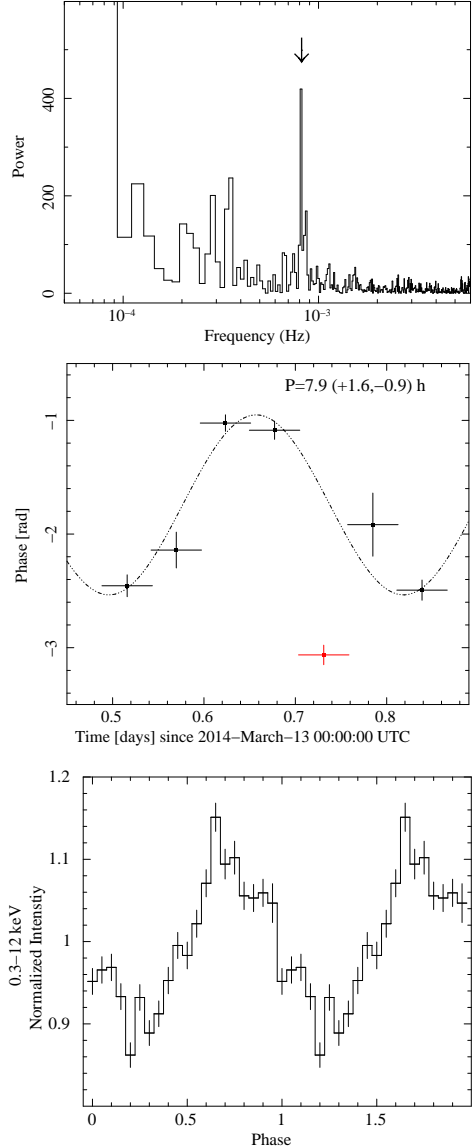

\begin{center}
\begin{tabular}{c}
\includegraphics[width=1.8in,angle=-90]{04powspec.ps} \\
\includegraphics[width=2.0in,angle=-90]{04orbital.ps} \\ 
\includegraphics[width=2.0in,angle=-90]{04_pf.ps} \\
\end{tabular}
\caption{\textit{Up}: PN plus MOS1 and MOS2 0.3--12 keV power spectrum of \bb. The peak at $\sim8\times10^{-4}$ Hz corresponding to the WD spin period is highlighted with an arrow. \textit{Center}: Evolution of the phase of the signal corresponding to the WD spin period versus time. The measured sinusoidal period is also reported in the upper right corner. The red points do not follow the sinusoidal trend and it is not included in the fit. \textit{Down}: X-ray pulse profile in the energy interval 0.3--12 keV. The folding period is 1218.7 s. Two cycles are shown for plotting purposes.}
\label{fig:04timing}
\end{center}
\end{figure}

The 0.3--12 keV PN background subtracted light curve shows a factor of $\sim2.5$ variability on timescales of tens of minutes through all the pointing. There is no evidence for long term variability associated with the optical 5--6 h orbital period. The power spectrum of the combined PN plus MOSs 0.3--12 keV light curve shows a peak at $\sim8\times10^{-4}$ Hz (see Fig. \ref{fig:04timing}). The exact period of the main peak was determined by means of a phase-fitting technique \citep[see][for more details on this technique]{dallosso03}, and it is $P^{X}=1222.6\pm2.7$ s. All uncertainties are hereafter at $1\sigma$ confidence level. This is consistent within 1.5 sigma with the optical period reported by \cite{thorstensen13}. We do not detect significant power at higher harmonics. To further test that the period is the true spin period of the WD, we folded the light curve at twice the 1222.6 value without finding changes in the pulse in two consecutive cycles. We conclude that 1222.6 is the period of the fundamental harmonic and thus it is the WD spin period. 

We also searched for variations of the phase of the spin period with time and found a sinusoidal trend (see Fig. \ref{fig:04timing}).  We infer a period of $7.9\pm^{1.6}_{0.9}$ h, consistent within $2\sigma$ with the longest (6.2 h) spectroscopic period found by \cite{thorstensen13}. We stress that this period is consistent with the total duration of the pointing and that one phase-bin does not lie on the sinusoidal trend 
(we did not include it in the fit).

We also studied the dependence of the pulses with the energy interval by folding the background subtracted light curves at the more accurate period found by period found by \cite{thorstensen13}, $1218.7\pm0.5$ s. A sinusoidal fit to the 0.3--12 keV folded light curve gives a pulsed fraction (PF) of $10.0\pm0.5$ per cent. Here the PF is defined as: 
$PF = (A_{max}-A_{min})/(A_{max}+A_{min})$, where $A_{max}$ and $A_{min}$ are the maximum and minimum value of the sinusoid, respectively. In Fig. \ref{fig:04timing} we show the 0.3--12 keV pulse profile.
We found that the PF slightly decreases with the energy interval: $PF_{0.3-2}=12.0\pm0.8$ per cent, $PF_{2-3}=11.0\pm1.2$ per cent, $PF_{3-5}=9.8\pm1.2$ per cent, and $PF_{5-12}=7.3\pm1.3$ per cent. 
We also inspected the phase dependence of hardness ratios but did not find any significant variations.

\subsubsection{The X-ray spectral analysis}

\begin{figure*}
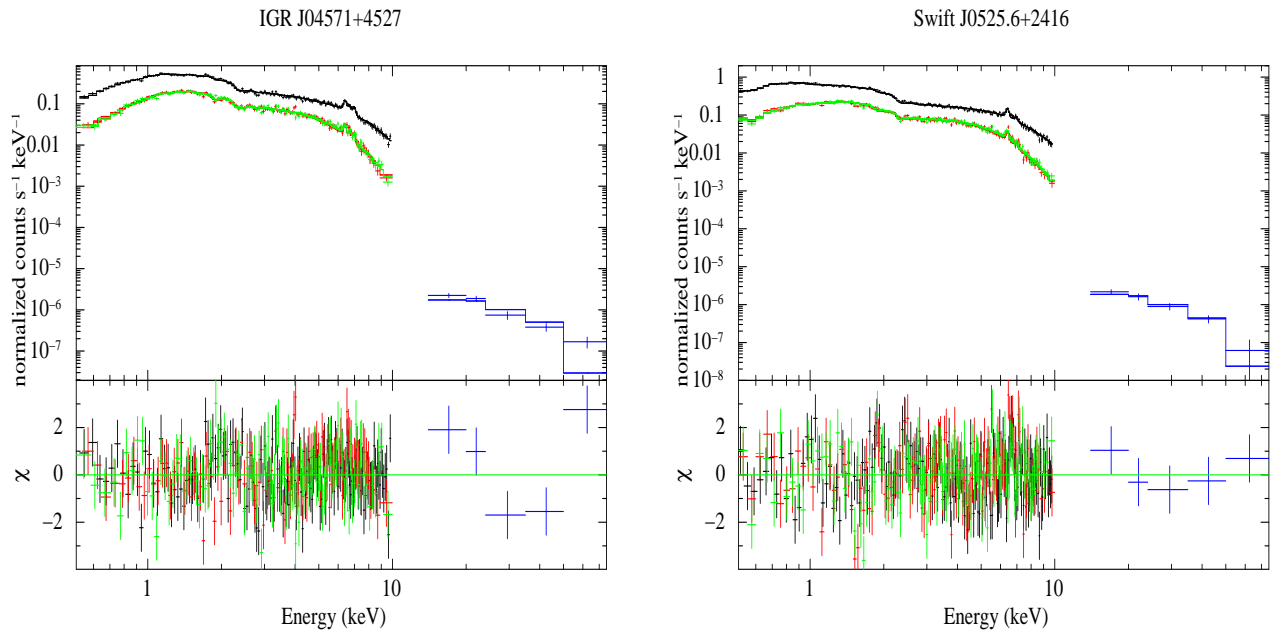

\begin{center}
\begin{tabular}{cc}
\includegraphics[width=3.3in, height=3.3in, angle=-90]{04spec.ps} & 
\includegraphics[width=3.3in, height=3.3in, angle=-90]{05spec.ps}\\ 
\end{tabular}
\caption{\textit{Left panel}: \bb\ broad-bad count spectrum. PN spectrum is in black, MOS 1 in red, MOS 2 in green and BAT in blue. Residual are shown in the lower panel. \textit{Right panel}: The same as the left panel, but for \aa.}
\label{fig:spec}
\end{center}
\end{figure*}

A single thermal model does not fit adequately the broad-band (0.3--70 keV) average spectrum. Instead, we found a satisfying spectral fit by using a model composed by the sum of two \textsc{mekal} (multi-temperature optically thin plasma) and a Gaussian at 6.4 keV, accounting for the fluorescent iron line. The spectral model also requires a complex absorption, made by a total \textsc{phabs} plus a partial covering \textsc{pcfabs} absorptions, as frequently found in magnetic CVs \citep[see e.g.][]{bernardini12}. The use of a partial absorber is dictated by the slight increase of PF at lower energies, suggestive of localised absorbing material. The results of the spectral fit are reported in Tab. \ref{tab:04_avspec} and shown in Fig. \ref{fig:spec}.
The hydrogen column density of the total absorber is lower than the total galactic column density in the direction of the source \citep[$5.4-6.5\times10^{21}$ cm$^{-2}$,][]{Dickey90,Kalberla05}, indicating an interstellar origin. 
The \nh\ of the partial (32 percent) covering absorber is about an order o magnitude larger, suggesting that it has a local origin. The  spectral fit indicates that the post-shock flow is dominated by rather high temperatures, without requiring a low-temperature plasma component \citep[see][and Sect. 3.2 of this work]{bernardini12}. We notice that a soft BB component is not required by the fits. 
We also fitted the broad-band spectrum with the model developed by \cite{suleimanov05}, which takes into account both temperature and gravity gradients within the post-shock region. This more physical  model, implemented in Xspec, allows a more reliable estimate of the shock temperature and thus the WD mass. To avoid the effects of the complex absorption we fitted the spectrum above 3 keV only and added two Gaussians lines that take into account the fluorescent iron line and the thermal complex. We derive $M_{\rm WD}=1.12\pm0.06\,M_{\odot}$ ($\chi^2_{\nu}=1.13$, 215 d.o.f.). This value is slightly larger but consistent, within uncertainty, with that derived using the hot \textsc{mekal} temperature ($1.03\pm^{0.05}_{0.20}\,M_{\odot}$).

\begin{table}
\caption{\bb\ spectral parameters for the 
best fitting model. We report the absorbed/(unabsorbed) $0.3-10$ keV flux (F$_{0.3-10}$), 
together with the $10-100$ keV flux (F$_{10-100}$).
Uncertainties are at the $1\sigma$ confidence level.}
\begin{center}
\begin{tabular}{ccc}
\hline 
\multicolumn{3}{c}{\bb\ average spectrum} \\
\hline 
N$_{H_{\rm P}}$             &  $10^{22}$ cm$^{-2}$ &  $0.302\pm007$   \\ 
N$_{H_{\rm pc}}$ 	        & $10^{22}$ cm$^{-2}$  & $5.4\pm1.0$        \\
cvf                         &  \%                  &  $32\pm3$          \\
kT$_{\rm hot}$              & keV                  & $63\pm_{19}^{8}$ \\
norm$_{\rm hot}$            & $10^{-3}$            & $5.8\pm0.4$        \\
kT$_{\rm cold}$             & keV                  & $6.8\pm^{1.7}_{1.2}$ \\
norm$_{\rm cold}$           & $10^{-3}$            & $1.3\pm0.4$        \\
A$_{\rm Z}$                 &                      &  $0.87\pm0.20$     \\
EW$^{a}$                    & keV                  &  $0.13\pm0.01$     \\
F$_{0.3-10}$                &  $10^{-11}$ \ergscm  & $0.973\pm0.010$  ($\sim1.3$) \\
F$_{10-100}$                &  $10^{-11}$ \ergscm  & $1.9\pm0.2$  \\
$\chi^2_{\nu}$ (dof)        &                      &  1.10 (362) \\
\hline 
\\
\end{tabular}  
\label{tab:04_avspec}                                                                                                         
\end{center}
$^{a}$ Gaussian energy fixed at 6.4 keV.\\
\end{table}
The spectra at pulse minimum ($\phi=0.12-0.37$) and maximum ($\phi=0.60-0.90$) were fitted using using the spectral model reported in Tab \ref{tab:04_avspec}, fixing the interstellar absorption and the abundance at their best fitting average value. We also fixed the hot \textsc{mekal} temperature to the average value, as it resulted to be unconstrained. This is not surprising given the weak PF at high energies. We get 
$\chi^2_{\nu}=1.01$, 274 d.o.f. and $\chi^2_{\nu}=1.03$, 246 d.o.f. for the pulse maximum and minimum respectively.
While all other parameters are constant within statistical uncertainties and consistent with the values found for the average spectrum, the only two parameters that change at maximum and minimum of the pulse are the $N_{H}$ of the partial covering absorber and the normalisation of the cold \textsc{mekal}. The former is $10\pm3\times10^{22}$ cm$^{-2}$ at minimum, whilst it decreases to $4.7\pm0.8\times10^{22}$ cm$^{-2}$ at maximum. The latter instead is $2.0\pm0.1\times10^{-3}$ at pulse maximum, while decreases to $0.7\pm0.3\times10^{-3}$ at minimum.

\subsection{\aa}

\subsubsection{The X-ray and optical timing analysis}
\label{subsub:05timing}

\begin{figure}
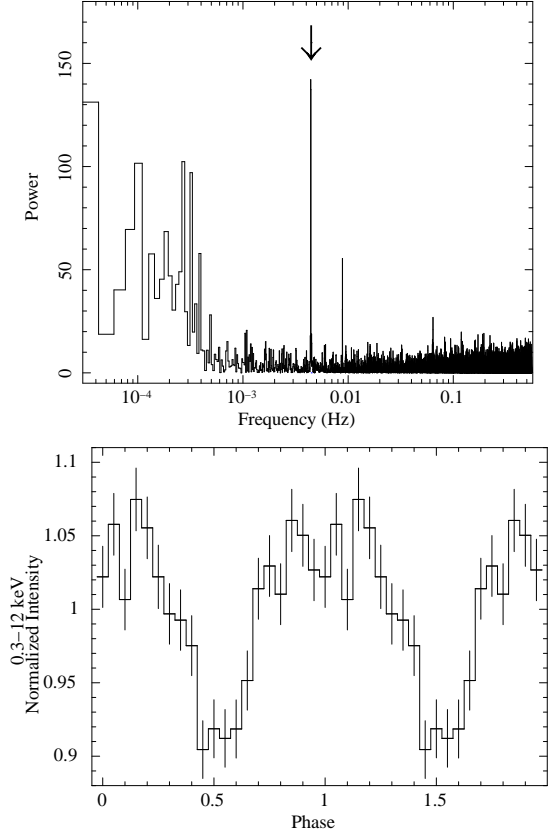

\begin{center}
\begin{tabular}{c}
\includegraphics[width=2.24in,angle=-90]{05powspec.ps} \\
\includegraphics[width=2.0in,angle=-90]{05_pf.ps} \\
\end{tabular}
\caption{\textit{Up}: PN plus MOS1 and MOS2 0.3--12 keV power spectrum of \aa. The narrow peak at $\sim4.4\times10^{-3}$ Hz, corresponding to the WD spin period, is highlighted with an arrow. The first harmonic is also evident. \textit{Down}: 0.3--12 keV pulse profile. The folding period is 226.28 s. Two cycles are shown for plotting purposes.}
\label{fig:05timing}
\end{center}
\end{figure}

We first visually inspected both the optical B band and the 0.3--12 keV PN background subtracted light curves using a small binning sample (10 s) to check for sign of the eclipse reported in \cite{ramsay09b} and we did not find any. While the X-ray light curve is showing short term variability, the B band one is almost flat with an average count rate of 4.8 c/s.
The power spectrum of the combined PN plus MOSs 0.3--12 keV light curve shows a narrow peak at $\sim4.5\times10^{-3}$ Hz (see Fig. \ref{fig:05timing}, central panel) with indication of a weaker harmonic at $\sim9\times10^{-3}$ Hz.
The period of the main peak, determined with the phase-fitting technique, is $P^{X}=226.28\pm0.07$ s. This is the first detection of a coherent signal from this source. There is no sign of orbital variability in the evolution of the phase of the main peak with time. Also in this case we checked that the 226.3 s period is the fundamental. 
We also do not find any long term trend ($>300$ s) in the X-ray light curve. Thus, we conclude that the spin period of the source is 226.3 s, and there is no evidence of eclipses.

The PF of the 0.3--12 keV light curve for the fundamental and first harmonic is $6.3\pm0.5$ per cent and $3.1\pm0.5$ per cent, respectively. In Fig. \ref{fig:05timing} we show the pulse profile for the 0.3--12 keV interval.
The PF slightly decreases as a function of the energy: $PF_{1-2}=10.3\pm1.1$ per cent, $PF_{2-3}=4.6\pm1.6$ per cent, $PF_{3-5}=4.7\pm1.4$ per cent, and $PF_{5-12}=5.4\pm1.4$ per cent. We did not include the 0.3--1 keV interval because of the low S/N. 
\noindent We also did not find any variation of the hardness ratios with phase.

\subsubsection{The X-ray spectral analysis}

The broad-band 0.3--70 keV spectrum requires more than one thermal component. We first used a model composed by the sum of two \textsc{mekal} and a Gaussian fixed at 6.4 keV absorbed by a total \textsc{phabs} plus a partial covering \textsc{pcfabs} absorber. We found residuals (in excess with respect to the model) at low energy around 0.55--0.65 keV, where OVIII and OVII lines are expected, respectively. 
We obtained a better fit by substituting the two \textsc{mekal} components with two \textsc{vmekal}, where the abundance of the different elements can be individually controlled. We left the oxygen abundance free to vary, linking the abundances of all other elements to a single, global, value and and found $\rm A_{O}=0.14\pm0.05$ compared to the Solar value. 
The abundances of elements other than oxygen, were found to be A$_{\rm Z}=0.52\pm0.20$\footnote{The abundances we derive depend critically on the abundance table we adopt. However, subsolar abundances are confirmed using both \cite{wilms00} and \cite{anders89} relative abundances, being the former a factor of about two larger. Since the abundances of O and Fe, the two elements that our data are most sensitive to, differ greatly between the two tables, this is not surprising.}. Summarizing, the \textsc{mekal} model predicts an overabundance of oxygen, this problem is resolved using the \textsc{vmekal} model that suggests a subsolar oxygen abundance. 
We also notice that a soft BB component is not required by the fits.
The $\chi^2_{\nu}$ is not satisfactory (1.29 for 395 d.o.f.), although this can happen when  dealing with PN and/or MOS data \citep[see e.g.][]{bernardini12} for this kind of sources. 
We have tried several alternative models but none resulted in an improved $\chi^{2}_{\nu}$. 
The results of the spectral fit are reported in Tab. \ref{tab:05_avspec} and shown in Fig. \ref{fig:spec}. 
We notice that the hydrogen column density of the total absorber is lower than the total galactic column density in the direction of the source \citep[$3.2-3.8\times10^{21}$ cm$^{−2}$][]{Dickey90,Kalberla05}. 
The temperature of the cool \textsc{mekal} is similar to those found in other IPs \citep[see][]{bernardini12}. 
The WD mass derived with the model presented in \cite{suleimanov05} is $M_{\rm WD}=1.01\pm0.06\,M_{\odot}$ ($\chi^2_{\nu}=1.18$, 231 d.o.f.). 
On the other hand, we get $M_{WD}=0.85\pm^{0.08}_{0.05}\,M_{\odot}$ when using the hot \textsc{vmekal} temperature. 
The lower values also reflect the use of only one-temperature (instead of a gradient of temperature) for the hottest post-shock regions.

\begin{table}
\caption{\aa\ spectral parameters for the 
best fitting model. We report the absorbed/(unabsorbed) $0.3-10$ keV flux (F$_{0.3-10}$) 
together with the $10-100$ keV flux (F$_{10-100}$).
Uncertainties are at the $1\sigma$ confidence level.}
\begin{center}
\begin{tabular}{ccc}
\hline 
\multicolumn{3}{c}{\aa\ average Spectrum} \\
\hline 
N$_{H_{\rm P}}$             & $10^{22}$ cm$^{-2}$  &  $0.31\pm0.02$   \\ 
N$_{H_{\rm pc}}$ 	        & $10^{22}$ cm$^{-2}$  &  $18.1\pm1.5$        \\
cvf                         &  \%                  &  $45\pm2$          \\
kT$_{\rm hot}$              & keV                  & $40\pm^{9}_{5}$ \\
norm$_{\rm hot}$            & $10^{-3}$            & $10.0\pm0.7$        \\
kT$_{\rm cold}$             & keV                  & $0.19\pm0.01$ \\
norm$_{\rm cold}$           & $10^{-3}$            & $6.0\pm_{2.4}^{3.5}$        \\
A$_{\rm Z}$ $^a$            &                      &  $0.52\pm0.20$     \\
A$_{\rm O}$ $^a$                &                      &  $0.14\pm0.05$        \\
EW$^{b}$                    & keV                  &  $0.12\pm0.03$     \\
F$_{0.3-10}$                &  $10^{-11}$ \ergscm  & $1.13\pm0.02$  ($\sim1.9$) \\
F$_{10-100}$                &  $10^{-11}$ \ergscm  & $1.9\pm0.2$  \\
$\chi^2_{\nu}$ (dof)        &                      &  1.29 (395) \\
\hline 
\\
\end{tabular}  
\label{tab:05_avspec}                                                                                                         
\end{center}
$^{a}$ Relative abundances from \cite{wilms00}.\\
$^{b}$ Gaussian energy fixed at 6.4 keV.\\
\end{table}

The spectrum at pulse minimum ($\phi=0.45-0.69$) and maximum ($\phi=0.85-1.05$) were fitted with the model presented in Tab. \ref{tab:05_avspec}, fixing the interstellar absorption and the abundance at their best fitting average value. We also fixed the hot \textsc{vmekal} temperature to the average value, as it resulted to be unconstrained. We obtain $\chi^2_{\nu}=1.15$, 282 d.o.f. and $\chi^2_{\nu}=1.12$, 289 d.o.f. for the pulse maximum and minimum, respectively. 
We found that the the density of the localized absorber is slightly lower at pulse minimum than at maximum, $14.6\pm2.0\times10^{22}$ cm$^{-2}$ with respect to $25.8\pm4.0\times10^{22}$ cm$^{-2}$, and that the normalisation of the hot \textsc{vmekal} is slightly higher at pulse maximum than minimum, $0.0164\pm0.0005$ with respect to $0.0092\pm0.0003$. The cold \textsc{vmekal} parameters and the covering fraction are found consistent, within uncertainties, with their average values.

\subsubsection{Optical spectroscopy}

\cite{ramsay09b} reported that three of the four WHT
spectra show Balmer series and HeII (4686 \AA) in emission. The second spectrum instead was much redder with no sign of emission lines, suggesting an eclipse of the source.
We investigated this in more details and found
that the second spectrum was offset in the spatial
direction by $\sim$4 pixels compared to the other three. 
It is likely that the telescope tracking was incorrect and thus another star, 4$^{''}$ away from the target, entered in the slit.
This star is not recorded in the USNO A2 or 2MASS
catalogues, but is detected in the WISE images and has a similar
brightness to \aa. This would explain the redder spectra and the absence of the emission lines. Thus, we conclude that \aa\ is not an eclipsing system.

\begin{figure}
\begin{center}
\includegraphics[width=2.3in,angle=90]{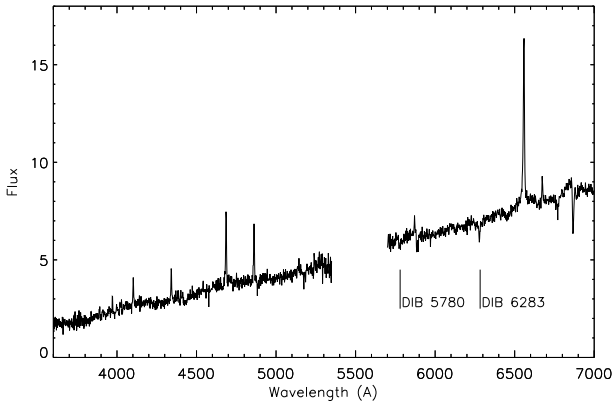}
\caption{The spectrum of \aa\ derived by combining three spectra
obtained using WHT and ISIS on the night of 2008 Oct 6. The spectrum
has been calibrated with a standard star to remove the instrumental
response, but is not on an absolute scale. The Balmer series of
Hydrogen are seen in emission as is He II (4686 \AA). 
H$_{\alpha}$ (6562.8 \AA) is the stronger line in emission.  
Vertical lines mark the position of diffuse interstellar bands (DIBs) at 5780 \AA 
(marginal detection) and 6283 \AA.}
\label{fig:dibs}
\end{center}
\end{figure}

We also searched for radial velocity variations in the emission lines. Over
the duration of 32 min, H$_{\alpha}$ was found to vary between --163
and --212 km/s ($\pm$15 km/s), giving marginal evidence of a radial
velocity shift. The spectrum shows evidence of interstellar reddening with diffuse interstellar bands (DIBs, see Fig. \ref{fig:dibs}) at 6283 \AA\ and possibly at 5780 \AA. We then used the combined spectrum to measure the equivalent width of the DIB feature at 6283 \AA\ as 1.0$\pm$0.1 \AA\ implying a colour excess $E_{B-V}$ in the range 0.39--1.31 using the relationship of \cite{cordiner11}. From the hydrogen column density of the total absorber derived from the X-ray spectral fits we derive E(B-V)$=0.44\pm0.04$ using \nh\,/E(B-V)$=6.8\times10^{21}$ \citep{ryter75}. We assume the latter one as the source colour excess.

The optical counterpart is also reported in the 2MASS catalogue as 2MASSJ05252270+2413332 with J$=14.65\pm0.07$, H$=14.16\pm0.08$, and K$=14.01\pm0.04$ magnitudes. 
It is also detected at longer wavelengths by the Wide-field Infrared Survey Explorer \citep[WISE;][]{wright10} as J052522.84+24133333.6 with W1 ($3.4\mu$) $=13.62\pm0.05$ and W2 ($4.6\mu$) $=13.49\pm0.06$.
The 2MASS J-H and H-K colour indexes corrected for intervening interstellar absorption are compatible with those of main sequence stars with spectral types in the ranges G8-K3 and G7-M1, respectively \citep{straizys09}.  
The WISE colours (W1 - W2)= 0.13$\pm$0.08 that can be considered free from intervening absorption imply a colour temperature of 4000$\pm$200 K compatible with a donor star in the range K5-M0. 
However, we must note that the accretion disc even in the nIR could mimic a late-type stars with different colours. Therefore, only a lower limit to the distance of about 600 pc can be given, assuming that the 2MASS K band absolute magnitude is that of a M1 star with $\rm M_K$ = 4.93 \citep{Knigge06}.

\section{Conclusions}
\label{sec:concl}

\bb\ shows a coherent X-ray signal with a period of $1222.6\pm2.7$ s, consistent with that found at optical wavelengths, confirming it as the WD spin period of the accreting WD. We found indication of an orbital modulation of about 8 h, which is close to the longer (6.2 h) period found in optical spectroscopy \citep{thorstensen13}.
The X-ray spectrum is well fitted with a multi-temperature optically thin emission with complex absorption as frequently found in magnetic CVs \citep[e.g.][]{bernardini12}. We conclude that \bb\ is an IP. 
Spectral changes with respect to the spin rotation of the WD indicates that the absorption originates from material in the pre-shock flow. The spin pulses are consistent with the generally accepted accretion-curtain geometry, where material flows towards the WD poles in an arc-shaped curtain \citep{rosen88}. When the primary pole points towards the observer the optical depth of the inflowing material above the shock is larger producing an X-ray flux minimum. 
We note that \cite{thorstensen13} estimate a high interstellar reddening for this source, E(B-V)$\sim$1.0, which implies N$_H\sim6.8\times10^{21}$ cm$^{-2}$. They suggested that this source is located farther than 500 pc as initially proposed by \cite{masetti10}. The value we found for \nh\ is a factor 2 lower and seems to favour the initial distance estimate of \cite{masetti10}. 
We derive a massive WD of 1.12 M$_{\odot}$, which is at the high end of CV WD masses \citep[see e.g.][]{ferrario15}.
 
\aa\ shows a coherent X-ray signal at 226.28 s that we identify as the WD spin period. Thus also this source is a CV of the IP type. There is no evidence for orbital variability in both the X-ray and optical \XMM\ light curves. From a re-analysis of previous optical spectra we do not confirm the presence of an eclipse. Further optical investigation is encouraged to constrain the orbital period.
The broad band X-ray spectrum of \aa\ can be fitted with a multi-temperature optically thin emission with complex absorption. Also in this source we find that the spin pulses are mainly due to photoelectric absorption originating in the pre-shock accretion flow, which is consistent with the accretion-curtain scenario. We found an indication of a subsolar abundance of oxygen. We derive a rather massive WD of 1.0 M$_{\odot}$. Although most hard X-ray IPs have been found to possess WD masses not so different from those of other CVs, this two new members appear instead to harbour rather heavy primaries. 

\section*{Acknowledgments}

This work is based on observations obtained with \XMM\ an ESA science mission directly funded by ESA Member States, with Swift, a NASA science mission with Italian participation. 
This publication also makes use of data products from the Two Micron All Sky Survey (2MASS), which is a joint project of the University of Massacusetts and the Infrared Processing and Analysis Center (California Institute of Technology), funded by NASA and National Science Fundation. 
It also makes use of data products from the Wide-field Infrared Survey Explorer (WISE), which is a joint project of the University of California, Los Angeles and the Jet Propulsion Laboratory/California Institute of Technology, funded by the National Aeronautics and Space Administration.
 
We acknowledge financial support from ASI INAF I/037/12/0

\bibliographystyle{mn2e}
\bibliography{biblio}

\vfill\eject
\end{document}